# Biologic and Prognostic Feature Scores from Whole-Slide Histology Images Using Deep Learning

Okyaz Eminaga [1,2,3,4], Yuri Tolkach [6], Rosalie Nolley [2], Christian Kunder [7],

Axel Semjonow [8], Martin Boegemann [8]

1) DeepMedicine.ai

2) Dept. of Urology, Stanford University School of Medicine

3) Dept. of Biomedical Data Science, Stanford University School of Medicine

4) Center for Artificial Intelligence in Medicine & Imaging, Stanford Medical School

5) Institute for Pathology and Cytology, Schuettorf, Germany

6) Dept. of Pathology, University Hospital Bonn, Germany

7) Dept. of Pathology, Stanford University School of Medicine

8) Prostate Center, Dept. of Urology, University Hospital Muenster, Germany

**Corresponding author:**

Okyaz Eminaga, M.D./Ph.D.

Department of Urology, Center for Artificial Intelligence in Medicine & Imaging (AIMI),

Laboratory of Quantitative Imaging and Artificial Intelligence (QIAI)

Stanford University School of Medicine

300 Pasteur Drive

Stanford CA 94305-5118

Tel: 650-725-5544

Fax: 650-723-0765

Email: okyaz.eminaga@stanford.edu

Stanford University, Stanford Medical School

# Abstract

Histopathology is a reflection of the molecular changes and provides prognostic phenotypes representing the disease progression. In this study, we introduced feature scores generated from hematoxylin and eosin histology images based on deep learning (DL) models developed for prostate pathology. We demonstrated that these feature scores were significantly prognostic for time to event endpoints (biochemical recurrence and cancer-specific survival) and had simultaneously molecular biologic associations to relevant genomic alterations and molecular subtypes using already trained DL models that were not exposed to the datasets of the current study. Further, we discussed the potential of such feature scores to improve the current tumor grading system and the challenges that are associated with tumor heterogeneity and the development of prognostic models from histology images. Our findings uncover the potential of feature scores from histology images as digital biomarkers in precision medicine and as an expanding utility for digital pathology.

## Introduction

Histopathology studies the morphological and structural changes of tissues caused by diseases and plays an essential role in tumor diagnosis and malignancy grading. The structural and morphological information from microenvironment and the epithelial components has shown to be essential for tumor definition and developing the malignancy grading system. A notable example for this is the Gleason grading system for prostate cancer. The Gleason grading system was first introduced by Gleason D.F., who initially defined five histopathological patterns according the microenvironment structures and epithelial components of the prostatic glands [1]. Gleason score is a well-validated prognostic parameter for survival outcomes and widely used in clinical decision making for prostate cancer cases [2,3]. Further, Gleason score has shown to be associated with genomic alterations frequently seen in prostate cancer [4,5]. Since its first introduction, the Gleason score has experienced several modifications in definition of Gleason patterns to improve the grading accuracy and the interobserver reliability [6]. Further, the Gleason score system was refined to the current Gleason grading system that includes 5 prognostic subgroups [7]. However, the Gleason grading system still inherits the major limitations of the previous versions of the Gleason score system that are associated with interobserver variation and criticized for not considering the nuclear morphology and the stroma component [3,8]. Therefore, other alternative grading systems like HELPAP score and the McKenney's malignancy grading system have been introduced to tackle these limitations; Helpap score is based on the glandular differentiation and the nuclear atypia of prostate cancer [9] and seems to improve the grading accuracy in low-grade prostate cancer [10]. McKenney et al introduced a complex histological grading system that considered the stroma patterns among other histological patterns [11]; his work revealed that the reactive stromal patterns in prostate cancer "stromagenic cancer" are significantly predictive for biochemical recurrence of prostate cancer[11]. From these observations, histopathology is primarily working on determining phenotypic features that are prognostic and/or a reflection of the disease biology. Therefore, understanding the development history of the current malignancy system is crucial to design and extract prognostic and biologically significant features from histology images using computational solutions. Several approaches are available to extract features from histology images including handcrafted feature extraction and the non-handcrafted feature extraction [12]. The deep learning is one of the advanced machine learning approaches and enables the construction of complex neural networks for solving the image classification problem [13]. The convolutional neural network is part of the deep learning and tackles the classification problems with multiple convolutional layers, where deep layers act as a set of non-handcrafted feature extractors that are quite generic and, to some extent, independent of any specific classification task [12,14]. Thousand features can be generated from images and utilized for regression analyses [15,16]. However, the internal validation of logistic or cox regression models is restricted by the number of events and the parameters being considered [17]. Harrel et al and Peduzzi et al suggested that the ratios of events per variable (EPV) should be at least 10 to maintain the validity of the logistic or cox regression model [18,19]. Given this, the results from previous studies showing lower EPV without any adjustment measurements for cofounders (e.g., the propensity score matching) are therefore limited [17,20].

The current study overcomes the limitations of previous works and provides novel feature scores that are prognostic and have biological significance. For the simplicity, the current study will evaluate the feature scores from Hematoxylin and Eosin (H&E) stained histology images that are related to prostate pathology. These feature scores are based on feature maps generated by customized convolutional neural network (CNN) models. The CNN models for prostate pathology were derived from PlexusNet[21] and Visual Geometry Group (VGG)[22], and developed on datasets independent from the dataset of the current study. The biological significance of feature scores was determined by identifying their associations with the genomic alterations described by the TCGA (The Cancer Genome Atlas) study for prostate cancer [4]. The logistic regression analyses for the relevant associations were validated by applying the 1000-samples bootstrapping. Further, we developed and internally validated the cox regression analyses for biochemical recurrence by meeting the EPV requirement and applying the 1000-samples bootstrapping on TCGA-PRAD dataset. The cox regression model was then tested on the McNeal's dataset that has a median follow-up of 9 years for cancer-specific survival to verify the prognostic values of these features on the unseen dataset for the unseen survival endpoint (i.e. prostate cancer-related death).

## Results

### Association between Feature scores and genomic alterations

After the feature scores were calculated from feature maps of the tumor lesions generated by detection models for high-grade prostatic intraepithelial neoplasia (HGPIN), Gleason pattern 3 (GP3) and 4 (GP4), cribriform (CR) and ductal morphology (DA), we tested their association with the genomic alterations, clusters and subtypes introduced by the TCGA study for prostate cancer [4]. For more clarification, we assigned each feature score to an alphabet to emphasize that these feature scores were calculated from different models (H: HGPIN model; G: GP3 model; P: GP4 model, D: DA model; C: Cribriform model). These feature maps or scores don't represent the findings for which the models were trained and are results of feature extraction process in the deep layers. **Supplement file 1** provides the cohort characteristic for genomic evaluation. **Supplement file 2** lists the screening results of the data distribution for feature scores and Gleason scores in 43 genomic alterations as well as the 5 subtypes and clusters mentioned in the TCGA study [4]. Only significant feature scores were considered for an associative analysis in each genomic variable. **Figures 1** and **3** illustrate results of histogenomic analyses that show the feature scores in two patients groups having distinct genomic alterations. **Supplement Table 1** provides the results of descriptive analyses for H, feature scores and Gleason grading (GG) after stratifying by associative genomic alterations (i.e., AR splice variant 7 status, the presence of TP53 hetero loss, and the presence of the high SPINK1 expression). The Gleason score or grading was determined at case level by pathologists from the TCGA study.

**Supplement Table 2** provides the results of univariate and multivariate regression analyses for features scores (H,G,D, P and C), Gleason scores, and their associated genomic alterations (i.e., AR protein expression level, AR score, AR RNA expression level, AR splice variant 7, fraction of altered genome, the presence status of TP53 heterozygous loss , and the existence of high SPINK1 expression). The following sections will present the genomic alterations and subtypes that were associated with feature scores. Since the H feature score was significant for many genomic alterations, we performed the t-Distributed Stochastic Neighbor Embedding (t-SNE) to visualize the distribution of these feature scores in connection with the related H&E histology images (**Figure 2**). Thereafter, we clustered the feature scores into 2 clusters using the K-mean clustering. Here, we evaluated 28 indices for clustering and applied the majority vote approach to determine the optimal cluster number for the feature scores using the R package "NbClust" [23].

Then, we evaluated the common prostate cancer morphologies of the histology images that are located in the centroid area of these clusters (diameter: 0.5 unit) and provided representative spots form these images in **Figure 2**. Finally, we evaluated the frequency of AR SV7 status in these clusters using the Fisher Exact test which showed that the distribution of AR SV7 status is significantly different between these clusters.

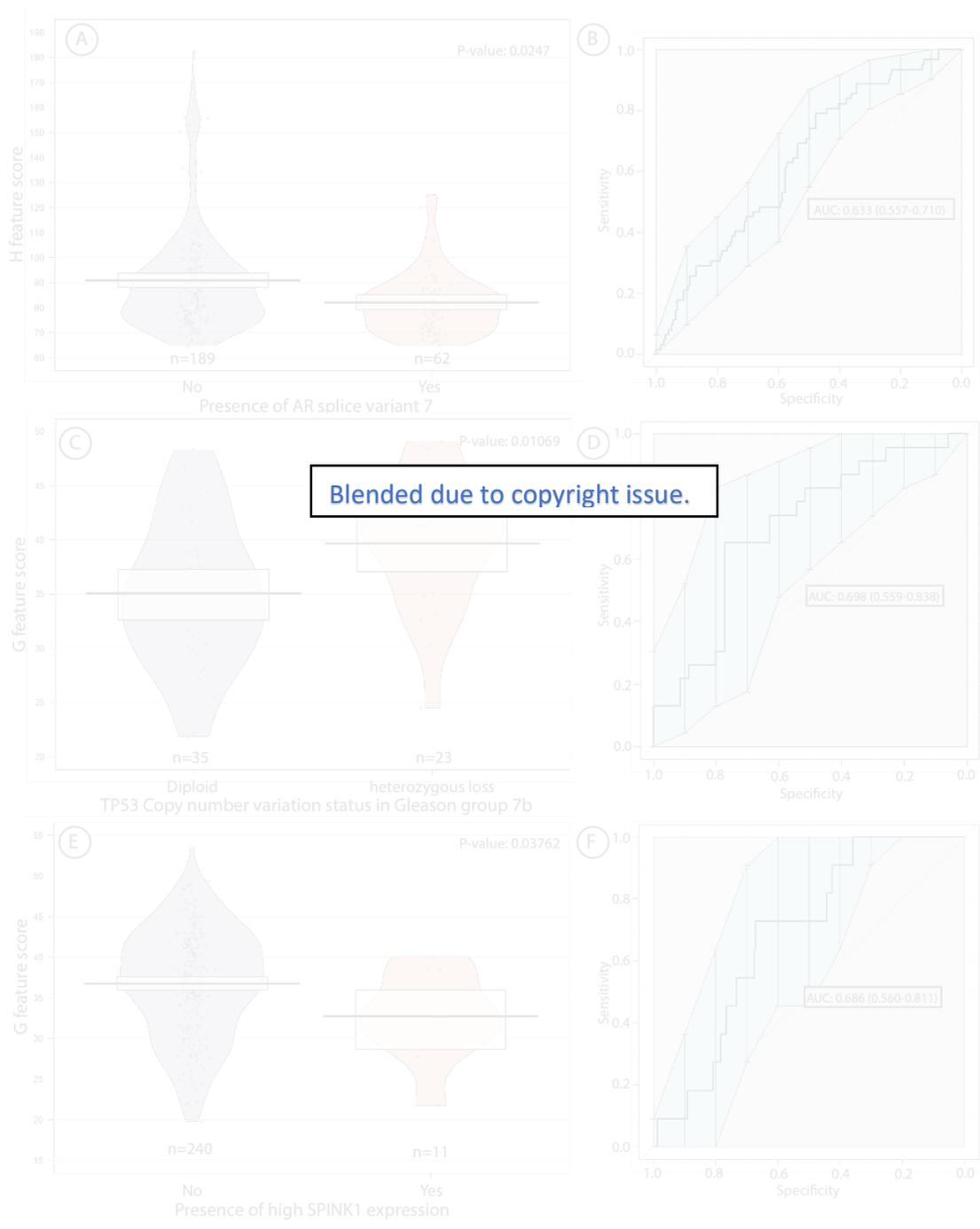

Figure 1: The pirates plots and the diagrams for the area under curve (AUC) of the receiver operating characteristic curve for features scores generated by models trained for the detection of HGPIN -H feature score- (A,B) and GP3 (C-F) -G feature score- and their associated genomic alterations. On the pirate plots (A,C,E), the middle line represents the mean value, the box represents the 95% confidence interval, the the colored areas represent the data density. P-values are estimated using the Wilcoxon–Mann–Whitney test. On the AUC-ROC diagrams, the overall 95% confidence intervals (CI) are given in parentheses. The 95% confidence of AUC was calculated by DeLong approach. The turquoise area represents the bootstrapped 95% CI of the sensitivity curves. With the exception of SPINK1 (10-times resampling), the 95% CI was calculated after applying resampling 1000-times.

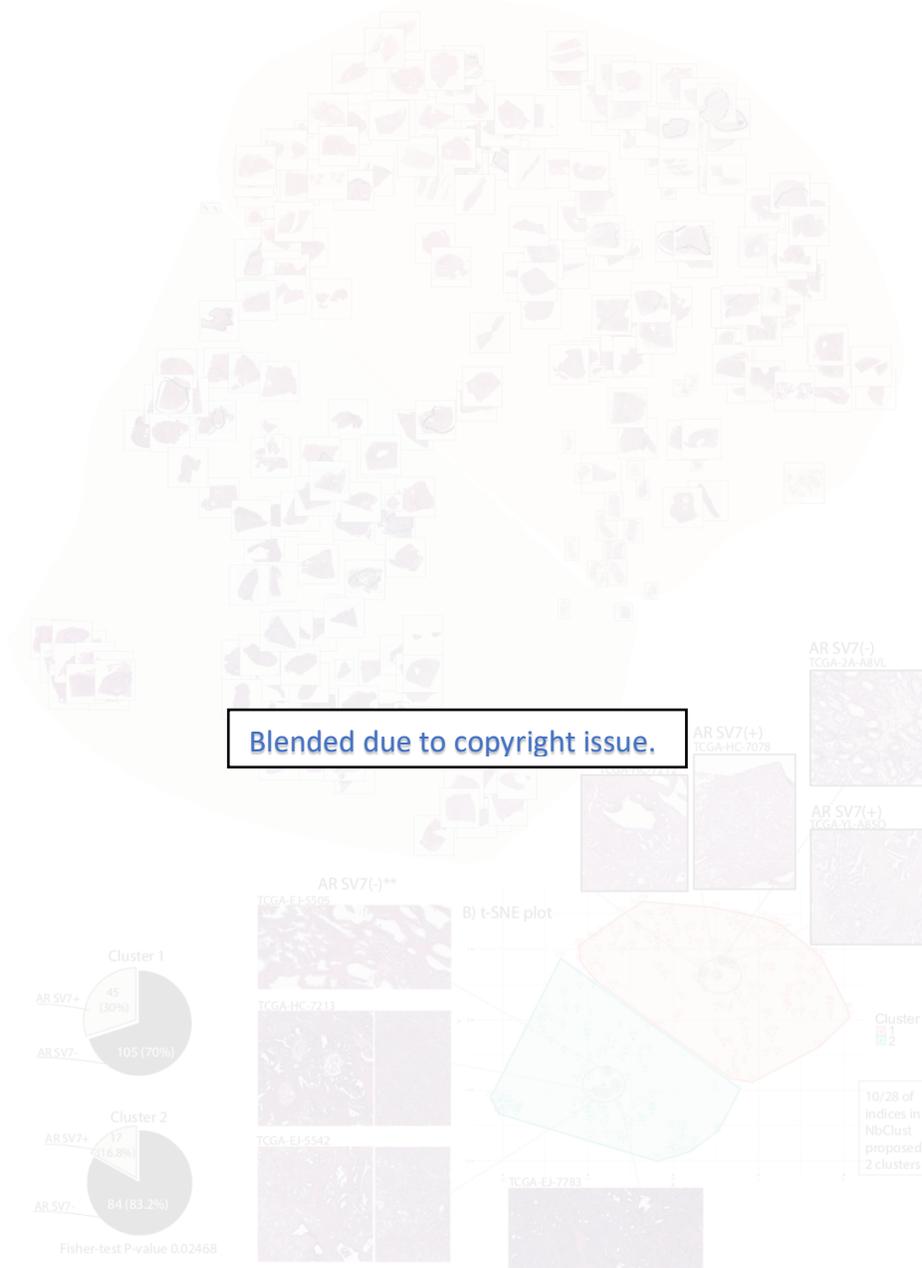

Figure 2 illustrates A) The data visualization of feature scores with a dimension of 1x160 generated by the detection model for HGPIN (H feature score) using T-distributed Stochastic Neighbor Embedding (t-SNE) in connection with the corresponding H&E-stained histology images. The gray circle defines the centroid area (Diameter: 0.5 units). (B) provides representative H&E histology images of common prostate cancer morphologies found on some H&E slides, whose feature scores are located inside the gray circle of each cluster (centroid area). ** denote that all selected cases from cluster 2 have a negative Androgen receptor splice variant-7 (AR-SV7) status. A high-resolution version of each snapshot in at 10x is available in the supplement section.

## Androgen-receptor expression

The H feature score and the Gleason score were independent predictors for the protein level of the Androgen receptor according to the multivariate analysis. The increasing H feature score was associated with decrease in the protein level of the androgen receptor [Odds ratio (OR): 0.998; 95% Confidence Interval (CI): 0.997– 0.999; AIC: -165.34; P=4.13e-05; Spearman correlation Coefficient (CC): -0.294]. Furthermore, the H feature score was an independent predictor for the derived numeric feature "AR score" (OR: 0.926; 95% CI: 0.882 – 0.974; AIC: 1284, P= 0.00191; Spearman CC: -0.2255), while Gleason score was not (P= 0.71337). In multivariate analysis including Gleason score, the G feature score was, in addition to Gleason score, an independent predictor for rise in the RNA expression level of the Androgen receptor (OR: 1.004; 95% CI 1.001 - 1.007; AIC: -171.15; P= 0.02525, Spearman CC: 0.222).

## Androgen-receptor splice variant 7

The H feature score is predictive for the presence of Androgen-receptor splice variant 7 (AR SV7) (OR: 0.967; 95% CI: 0.944– 0.987; P=0.00305, AIC: 272.8].  In a multivariate regression analysis including Gleason score and H feature score, the H feature score was an independent predictor for the presence of AR SV7 (P=0.0032), whereas the Gleason score was not (P=0.5733); the mean H feature score of AR SV7 cases was 82.1 (95% CI: 79.3 - 85.2) and the cases with absence of AR SV7 had a mean H feature score of 90.9 (95% CI: 88.0 - 93.8) (**Figure 1A**). The AUROC (area under received characteristic curve) of the H feature score for the AR SV7 status prediction was 0.633 (95% CI: 0.557-0.710) as shown in **Figure 1B.**

## Fraction of altered genome

In addition to Gleason score, the D and H feature scores were independent predictors of the fraction of altered genome (D: P=0.0281, H: P=0.00159) while the G feature score is not (P=0.52583) as shown in **Supplement Table 2**. The spearman correlation coefficient (CC) for the D feature score was -0.16 (P=0.01305) and for the H feature score was 0.14 (P=0.03), while the Gleason score had the strongest correlation with the fraction of altered genome (Spearman CC: 0.50; P=4.32e-15).

## TP53 heterozygous loss

The multivariate analysis revealed that Gleason score was an independent predictor for the presence of heterozygous loss (hetero-loss) of the TP53 (P=9.19e-05) whereas the G feature score was not predictive for TP53 hetero-loss (P= 0.184). By evaluating the G feature scores by the TP53 hetero-loss status, the mean G feature score was 38.0 (95% CI: 36.6 - 39.3) for cases with the TP53 hetero-loss and 35.9 (95% CI: 34.9 - 37.0) for cases with diploid TP35 gene. By stratifying the G feature score according to GG groups, the G feature score was a significant predictor for the TP53 hetero-loss only in "the 4+3" GG group [OR: 1.119 (95% CI: 1.027 - 1.234), P= 0.01537, AIC: 75] with AUROC of 0.698 (0.598 – 0.838). In cases with the Gleason grade 7b (4+3), the mean score of the GP3 features was 39.7 (95% CI: 37.2 - 42.2) for the TP53 hetero loss and 35.1 (95% CI: 33.0 - 37.2) for the TP53 diploid.

## SPINK1 Expression

Multivariate regression analysis including the G feature score and Gleason score showed that the G feature score was the only predictive parameter for elevated SPINK1 expression. An increase in the G feature score was associated with decreasing probability of the presence of elevated

SPINK1 expression (OR: 0.906, 95% CI: 0.828-0.978; P=0.0358; AIC: 91; AUROC: 0.686, 95% CI: 0.560-0.811). The mean G feature score was 32.7 (95% CI: 29.4 - 35.7) for cases with high SPINK1 expression and 36.8 (95% CI: 35.8 - 37.6) for cases with non-elevated SPINK1 expression. Due to the low number of 11 cases with SPINK1 overexpression, we applied 10 times bootstrapping.

## SPOP mutation

Given that a recent study found that a SPOP mutation can be detected on the basis of histology images in prostate cancer [24], we conducted an association analysis and were able to identify one of our feature scores (P feature score) was associated with the SPOP mutation after applying a rectifier that considers the negative values and then inverts these values to positive; the transformed P feature score was an independent predictor for SPOP mutation (OR: 0.991, 95% CI: 0.983-0.998; P=0.0148; AUROC: 0.625, 95% CI: 0.519-0.731; AIC: 177.35) in multivariate analysis including Gleason grade. Twenty-eight cases with the SPOP mutation had a mean value of 85.2 (95% CI: 66.1-104.3), whereas 223 cases with SPOP negative mutation had a mean value of 117.2 (95% CI: 107.4-127.0).

## mRNA clusters

The TCGA study determined 3 clusters for mRNA expression profiles and we identified an association between these clusters and the H and P feature scores. The post-hoc analyses reveal that the Gleason scores were significantly higher in cluster 2 compared to cluster 1 (adj. P= 0.0094657) and 3 (adj. P= 0.0004570). The P feature score in cluster 3 differs significantly from that in cluster 1 (adj. P=0.0259211) while no significant difference in the P feature score was observed between clusters 2 and 3 (adj. P= 0.141) or clusters 2 and 1 (adj. P= 0.786). In addition, the H feature scores were significantly different between cluster 2 and 3 (adj. P= 0.041) and between 1 and 3 (adj. P= 0.0690949), although the post-hoc analysis could not determine any significant differences in H feature scores between cluster 1 and 2 (adj. P= 0.978). Figure 3 summarizes the post-hoc analyses between the feature scores, Gleason score and mRNA clusters.

## miRNA clusters

Six clusters for the miRNA expression profiles were identified by the TCGA study. The post-hoc analyses revealed that the Gleason scores were significantly lower in clusters 1 and 2 compared to clusters 3-6 (adj. P range between 4.680e-4 and 0.0299005) with an exception between clusters 1 and 4 (adj. P= 0.089767) as shown In **Figure 3**. The P feature score was significantly lower in cluster 6 compared to clusters 2, 4, and 5 (adj. P range between 0.0077803 and 0.0221771). The average P feature score in cluster 6 was comparable to that in cluster 3 (adj. P=0.3539039). Cluster 4 had significantly a lower H feature score compared to cluster 5 (adj. P= 0.093) and 6 (adj. P= 0.031), while the G feature scores were comparable across all clusters (adj. P range between 0.1651872 and 0.9999994).

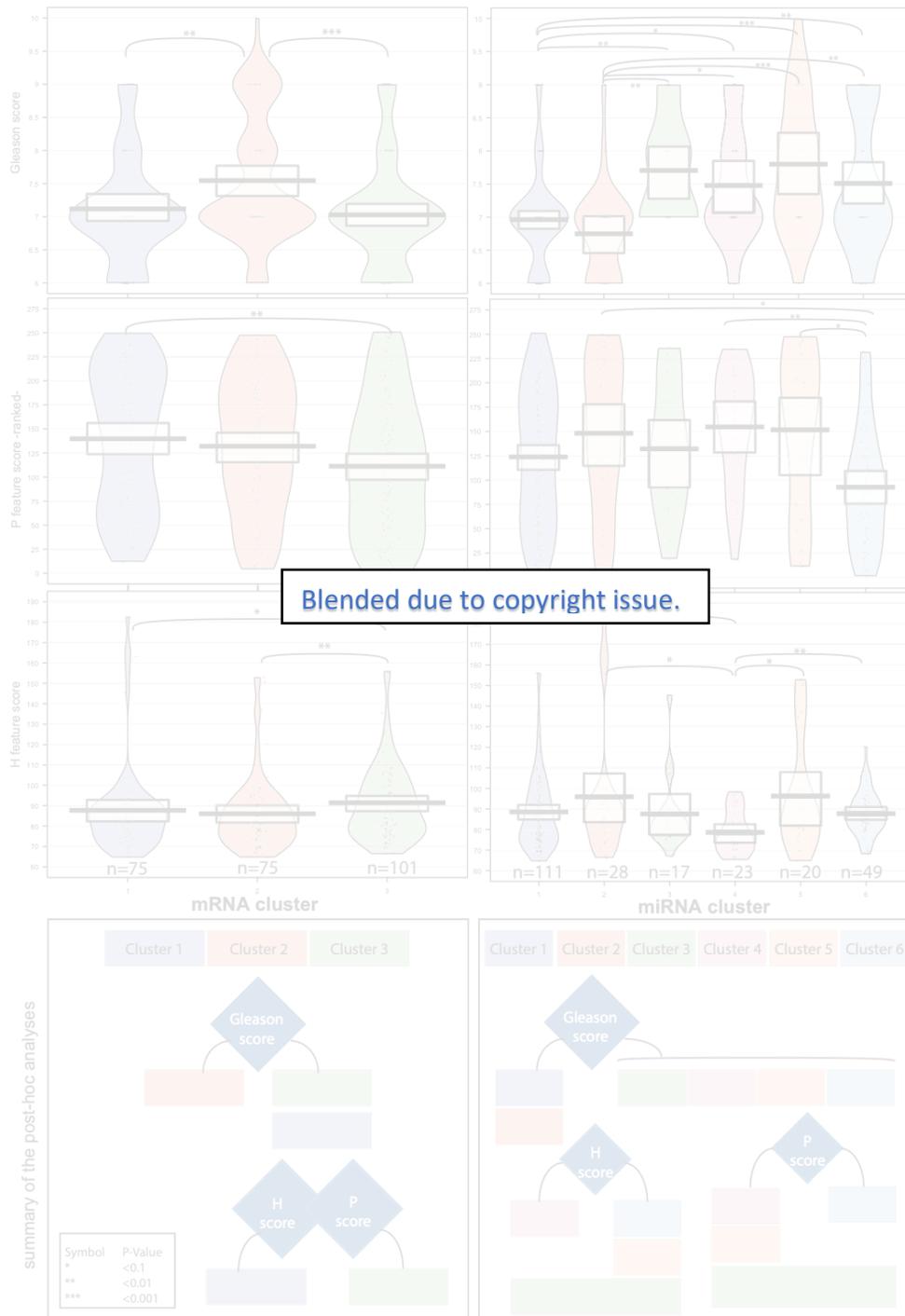

Figure 3: An illustrative representation of post-hoc analyses results according to Tukey that evaluated the variation in P and H feature scores and Gleason score between miRNA and mRNA clusters defined by the TCGA-PRAD study. A summary of the post-hoc analysis is also provided for miRNA and mRNA clusters. Adj. P-value: Adjusted P-value according to Tukey.

## Feature scores as prognostic factors

Patient characteristics for TCGA and McNeal cohorts are provided in **supplement Table 3.** Seventy-eight cases were excluded from TCGA-PRAD cohorts because of missing follow-up data. The median follow-up time for the TCGA-PRAD dataset was 2.57 years whereas the McNeal dataset had a median follow-up time of 9 years. In TCGA-PRAD dataset, forty-six patients (12.7%) had a biochemical recurrence, whereas 5 patients were died during the 9 years follow-up in the McNeal dataset. According to the univariate cox regression analyses, G and C feature scores were predictive for biochemical recurrence (BCR) whereas H, P and D were not; when all feature scores included in a multivariate cox regression analysis, only H and G were independently predictive for BCR (**Table 1**). By applying the stepwise cox regression analysis, we found that a model containing H, G and D feature scores provided the most suitable goodness-of-fit (AIC = 472) and were therefore considered for the multivariate cox regression analyses. In multivariate analyses, Gleason grading and the G feature score were independent prognostic factors for BCR whereas the H and D feature scores were not. We repeated the multivariate analyses after replacing the Gleason grading with the tumor stage (pT) and found that the G and H feature scores were independent prognostic factors for BCR (**Table 1**). The likelihood ratio test statistic (distributed chi-squared: 28.095, P=1.155e-07) revealed that a model with these three feature scores fits significantly better than a model containing the Gleason grading group only.

We built two cox regression models for BCR prognosis (**Table 1**) to identify the effect of combining GG groups and these feature scores; the first cox regression model considered the three feature scores mentioned above; the second cox regression model included, in addition to the previous feature scores, the GG groups. The first model achieved a c-index of 0.706 (95% CI: 0.606-0.779). The second cox regression model combining Gleason grading and the feature scores achieved a higher c-index of 0.748 (95% CI: 0.720-0.821), emphasizing that this combination improves the discriminative performance for BCR prognosis. The Schoenfeld residuals was applied to check the proportional hazards assumption and showed that the ratio of the hazards for all three feature scores is constant over time (**Supplement file 1**). The linear predictors from each model was binarily categorized into two risk groups by applying the mean value of 0.61 as threshold to estimate the survival function using the Kaplan-Meier Curves. The KM curve from **Figure 4A** showed that these risk groups determined by the first cox regression model -model A- are significantly discriminative (Log-rank P-value<0.0001). In one hand, we found that the low-risk group (n=177) achieved a 2-year BCR-free survival rate (PFS) of 97.4 % (95% CI: 94.9 - 100.0%) and a 5-year PFS of 88.3 % (95% CI: 80.6 - 96.7%). In another hand, the high-risk group (n=194) showed a 2-year PFS of 86.3 % (95% CI: 81.1 - 91.8%) and a 5-year PFS of 66.9 % (95% CI: 54.6 - 0.82.1%). When the second cox regression model (model A + GG groups) -model B- was applied to BCR prognosis, the gap between the two curves for the low-risk and high-risk groups was wider emphasizing that the  both groups from model B are more discriminative than those from model A (**Figure 4B**). Additionally, we validated the cox regression model that predicted the cancer-specific survival of 68 McNeal cases who underwent radical prostatectomy after adjusting the 5-year cancer-specific survival rate of the cohort close to 94% according to Wight et al [25]. After stratifying these McNeal cases into to the binary risk groups, the Kaplan-Meier Curve showed that these risk groups (Log-rank P-value= 0.0025) are also discriminative for cancer-specific survival, although these risk groups were developed for the prognosis of the biochemical recurrence as shown in **Figure 5**. Based on the risk groups defined by model A, no cancer-specific death event was registered in the low-risk group (n=43) during a 9-year follow-up after the surgical treatment [5-year cancer-specific survival (CSS) rate: 100%], while the high-risk group (n=25) showed a 5-year CSS rate of 83.6% (95% CI: 70.2-99.7%)]. According to the risk groups defined by Model B, the low-risk group (n=35) also showed a similar cancer-specific survival rate of 100% during the 5-years follow-up after the surgical treatment (**Figure 5A**) while the high-risk group (n=30)  had a 5-year CSS rate of 87.6% (95% CI: 76.9-99.8%) (**Figure 5B**). To note that the Gleason grading 7a or 7b didn't show any significant differences in survival outcome on McNeal dataset. **Supplement file 3** includes Kaplan Meier Curves for Gleason grading groups on TCGA datasets and the adjusted McNeal dataset. It also provides Kaplan Meier Curves for the risk groups and Gleason scores for the larger dataset from McNeal dataset (n=125) which reveal that the results did not change even after including additional cases.

*Table 1: The univariate and multivariate cox regression analyses. * denote the that that 95% confidence interval was calculated by the bias-corrected and accelerated bootstrap method. HR: Hazard ratio; C-index: Concordance index. pT: pathological AJCC tumor stage; AIC: Akaike information criterion; EPV: Events per variable.*

Blended due to copyright issue.

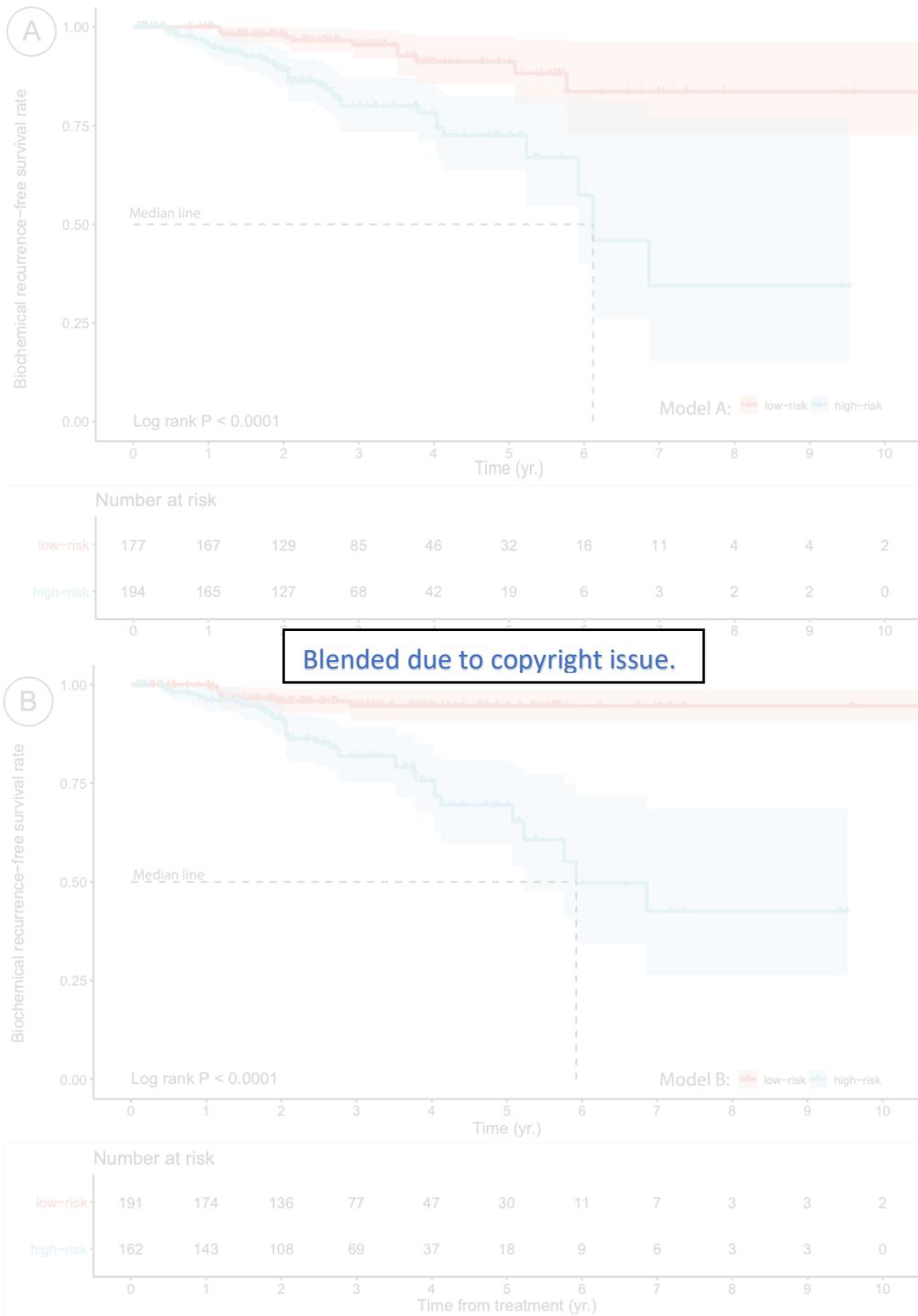

*Figure* 4 *The Kaplan Meier curves for biochemical recurrence-free survival for model A (H, G and D feature scores) and model B (H,G and D feature scores, and Gleason grading groups).*

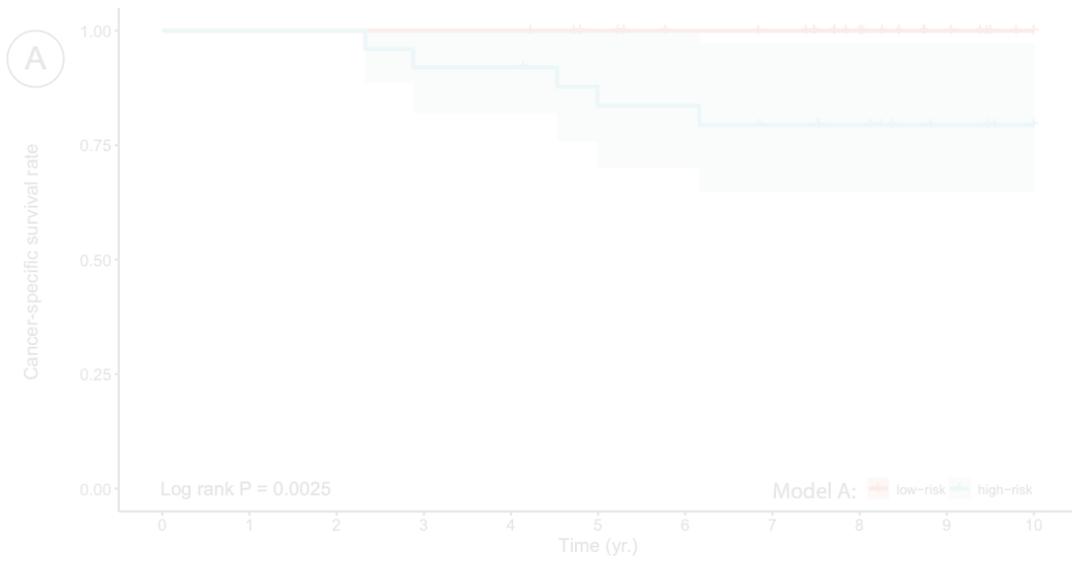

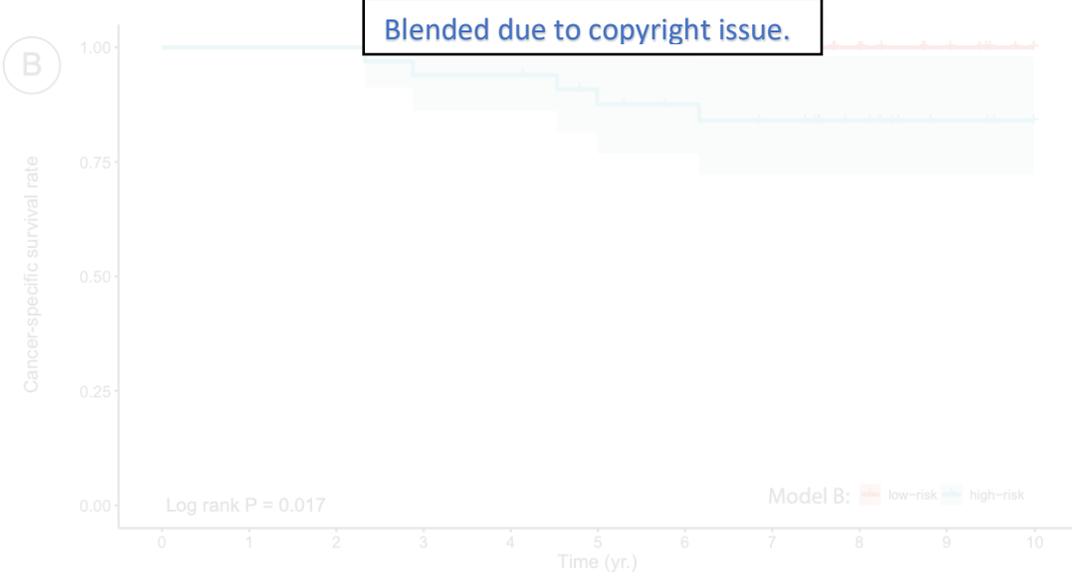

*Figure 5: The Kaplan Meier curves for 68 cases from the McNeal data set where the cancer-specific survival rate of the cohort was adjusted to 94% 5-year cancer-specific survival rate according to the data from Wight et al [25].*

## Discussion

We provide for the first-time feature scores from histology images that are prognostic and have simultaneously biological associations. The discovery of these feature scores is confirmed by the well-established fact that histopathology is a reflection of cancer biology and tumor progression and histopathology is prognostic for the outcome. Further, the current study shows that DL models initially developed to determine the tumor precursor and tumor grading generate feature maps with relevant information for prognostic and biological associations from "unseen" datasets. Our rigorous validation of the feature scores for their prognostic values reveal that these feature scores are robust prognostic feature scores even after changing the endpoint from the biochemical recurrence (BCR) to cancer-related death. According to the model comparison analysis, a cox regression model that combines three feature scores is better fitted to outcome prediction than the Gleason grading system. Further, the incorporation of feature scores into the Gleason grading system enhances its discriminative accuracy (c-index improvement from 0.706 to 0.746), highlighting the potential to improve the current Gleason grading system with these feature scores. According to the literatures, a previous work showed the feasibility to determine specific mutations based on histology images in lung cancer [26]. One other study introduced the application of genomic data and histology images to predict the outcome in prostate cancer[27]. However, none of the previous studies introduced feature scores from histology images that are prognostic and biologically significant at the same time. Moreover, none of the previous studies was able to provide such robust prognostic features for prostate cancer.

Our study evaluated the association between the feature scores and the most relevant genomic alterations related to prostate cancers. Here, we found that the feature scores are significantly correlated with the androgen receptor at different levels (protein, androgen score, mRNA expression and splice variant 7), SPINK1 overexpression and the heterozygous loss of TP53, the presence of SPOP mutation and weakly associated with the fraction of altered genome, emphasizing that these feature scores are potential digital biomarkers for genomic alterations and molecular subtypes of prostate cancer.

The androgen activity plays a crucial role in prostate cancer pathogenesis and responsible for the creation and overexpression of most ETS (erythroblastosis virus E26 transformation-specific) fusions in prostate cancer [28,29]. Gene fusions involving ETS family transcription factors are found in approximately 50% of prostate cancers and are the basis for the molecular subclassification of prostate cancer[4,30,31]. Additionally, the mRNA and miRNA clusters identified by the TCGA research team for prostate cancer have shown to be tightly correlated with ETS fusion status[4]. The androgen receptor (AR) is a ligand-dependent transcription activator for the androgenic hormones like testosterone or the more potent dihydrotestosterone [32-34]. AR regulates the normal prostate development and prostate function[35-37], as well as the growth and the progression of PCa[33]. Additionally, AR is highly expressed in androgen-dependent and recurrent prostate cancer[38]. The AR splice variant-7 (AR-SV7) is a truncated form of the androgen receptor that lacks the ligand-binding domain and is linked to the development of the castration-resistant prostate cancer [39,40]. The truncated variants of AR are capable of activating AR target genes in the absence of androgens [41]. Further, the AR-SV7 overexpression in primary prostate cancer has been associated with a worse prognosis after radical prostatectomy[42]. SPINK1 is serine protease inhibitor Kazal-type 1 and generally overexpressed in ETS-fusion-negative PCA and hence represent a molecular subtype of PCa[4]. The overexpression of SPINK1 stimulates the cell proliferation and invasion in ETS-fusion–negative PCa[43,44] and associated with poor prognosis [45,46]. The tumor protein p53 (TP53) is one of most studied genes or proteins in cancer research and regulates the cell cycle and hence functions as a tumor suppression [47]. The alteration of TP53 leads to compromises cellular cycles inducing apoptosis in DNA damaged cells and consequently promotes tumor progression through acquisition of additional genetic alterations[48]. Further, the TP53 copy number loss in primary tumor of prostate cancer was found to be prognostic for recurrence and metastatic events [49]. The SPOP gene is another tumor suppressive gene in prostate cancer that encodes for the substrate-recognition component of a Cullin3-based E3-ubiquitin ligase[50]. Mutations in SPOP represents the most common point mutations in primary prostate cancer (6-13%) [4,51], where SPOP-mutated PCa shows the highest transcriptional activity of AR[52]. Further, SPOP mutation drives the tumorigenesis of prostate cancer by affecting the both PI3K/mTOR and AR signaling [53]. Although the prognostic value of SPOP mutation is unclear, its expression level was associated with poor outcome in primary prostate cancer [54].

The current study has some limitations that warrant mention. First, the current study used a prospectively collected retrospective data and therefore inherits the limitations of a retrospective study. We aim to initiate a prospective study that would confirm the clinical utilization of these feature scores for prognosis and detecting the subtypes. Second, we didn't consider the whole slides of each case and therefore the tumor heterogeneity that may have impact the outcome. However, the current study considered using histology images from the index lesions to confirm the prognostic value of the feature scores (The largest tumor lesion in the prostate) as the tumor grade of the index lesions significantly influences the final tumor grade of the patients and has shown to be associated with survival outcome [55,56].

In summary, the current study identified novel feature scores from histology images that have associations with molecular subtypes of prostate cancer and show prognostic values. Such feature scores can be utilized as digital biomarkers for personalized precision medicine in prostate cancer. Such digital biomarkers will moreover facilitate identifying subsets associated with worse outcomes or patient groups that may benefit from specific treatment regimens using a non-invasive computational algorithm. Further, these feature scores can potentially improve the current Gleason grading system as our findings reveal. Nevertheless, further detailed evaluation is required to validate the improvement of the Gleason grading system by these feature scores.

## Materials and Methods

For the current study, we examined only DL models that were not trained on the TCGA datasets in order to avoid any possible interaction or bias associated with training on the TCGA-PRAD dataset from the previous study that introduced different models for prostate pathology. For that reason, we considered feature matrices generated by detection models for HGPIN, Gleason patterns 3 and 4, and cribriform and ductal morphology. Detection models for inflammation, prostate cancer, vessel and nerve, and Gleason pattern 5 were trained on datasets having TCGA-PRAD images and therefore excluded from the current study. For each whole-slide H&E-stained diagnostic image, the feature matrices were derived from tumor areas and stored in the cMDX file. For simplicity, we will describe the calculation steps of a single feature score. First, we computed the median value of each column of the feature matrix, that cover all patch images from the tumor regions of a single WS image. Then, the column-wise median values were added to a single value called the feature score of the particular model that generated the feature matrix. The reason of considering the median value is that the median value is less affected or distorted by outliers. To simplify the description of the feature scores, we assigned each feature score to an alphabet to emphasize that these feature scores were calculated from different models (H: HGPIN model; G: GP3 model; P: GP4 model; D: DA model; C: Cribriform model). The reader should keep in mind that these feature matrices or scores don't represent the findings for which DL models were trained and are results of feature extraction process in the deep layers of these models.

After the feature maps were stored in the cMDX file format for all whole-slide images, the features scores were calculated and exported using custom tools for cMDX files as comma-separated values (csv) files. Later, we merged the clinicopathological information from the data hub of Xena browser (https://xenabrowser.net/hub/) and the csv file with features scores together by using the TCGA-PRAD case identification number. Finally, the genomic alterations, clusters, and subtypes introduced by the TCGA-PRAD study were considered for genomic analyses [4].

## Histogenomic Evaluation

We identified 251 diagnostic H&E whole-slide images from TCGA-PRAD with available results for genomic alterations mentioned by the TCGA article. We determined the importance of features extracted by our DL models by testing their association with the genomic alterations, clusters and subtypes introduced by the TCGA-PRAD study [4]. Here, we evaluated the data distribution of feature scores and Gleason score in 30 genomic alterations and 5 subtypes and clusters mentioned in the TCGA study by applying non-parametric tests such as the Mann-Whitney test for binary classification, the Kruskal-Wallis Test for parameters with multiple classes, and the Pearson correlation test for two continuous values. Genomic alterations, subtypes, or clusters whose score distributions across their subgroups were not statistically identical at the 95% significance level were selected. For genomic alterations with unequal Gleason score distributions, the feature scores were stratified according to Gleason grading groups. Pirate plots were applied to illustrate the differences between categorical and continuous parameters. The odds ratios of the presence of categorized genomic alterations was estimated for each incremental increase of feature score using the logistic regression algorithm. The Area-under-the curve (AUC) and the corresponding 95% Confidence interval was evaluated for each feature score by the binary category of the genomic alteration. For subtypes or clusters with more than 2 categories, we applied the Post-hoc test according to Tukey's Honestly Significant Difference after ranking the feature scores. A difference was considered significance if its adjusted two-sided p-value (adj. P) according to Tukey was below 0.1. For all tests, with the exception of the post-hoc and distribution analyses, the bootstrapping technique was applied to internal validation on 1,000 samples [57].

## Prognostic Evaluation

### Biochemical Recurrence-Free Survival

We identified 371 diagnostic FFPE H&E whole-slide images from TCGA-PRAD with follow-up for biochemical recurrence (BCR); BCR is defined by a consecutive rise of prostate-specific antigen - PSA in at least two measurements or administration of adjuvant therapy for evidence of detectable PSA >0.1 ng/ml for at least 6 weeks postoperatively[4,58].

Univariate and multivariate cox regression analyses for BCR compromised H, G, P, C, and D feature scores. The Akaike information criterion (AIC) assessed the relative goodness-of-fit of the Cox regression models. We then applied stepwise cox regression to identify predictive variables that fit the cox regression model with the best relative goodness-of-fit. The variables from a cox regression model with the lowest AIC were selected for further analysis. Predictive accuracy estimates are generally quantified with concordance index. Two multivariable Cox regression models were developed. The first cox regression model (Model A) included the selected variables a and was fitted for BCR prediction. The second cox regression model (Model B) contained the variables from Model A and the Gleason grading group. The linear predictor for BCR was estimated using these cox regression models. The linear predictor was then scaled between 0 and 1 to define the prognostic score. The mean prognostic score was selected as cutoff to stratify the cohort into two groups. Kaplan-Meier plots illustrated BCR-free survival rates after stratification, according to tumor stage (pT), Gleason grading (Gls) and the binary prognostic score were generated.

## Cancer-specific Survival

We applied the cox regression models for biochemical recurrence to predict cancer-specific survival (CSS) on two cohorts for external validation. The first cohort was adjusted to the 5-year CSS rate of 94% according to the survival data from Wright et al [25] and consists of 68 cases from McNeal's dataset with a median follow-up of 9.4 years. The second cohort contained 125 histology images from McNeal's dataset with a median follow-up of 10.0 years. These cases were randomly selected. We chose one slide representing the index lesion identified by being the largest lesion in the prostatectomy specimens according to a graphical diagram representing the tumor distribution from each case [59]. Further, we weighted cases with Gleason score 7a (3+4) and 7b (4+3) as cases with Gleason score 6 have a cancer-specific survival rate of 100% and it is clinically more important to stratify patients with higher Gleason scores according to their increased risk of cancer-associated death [60] and the CSS of Gleason score 7b was comparable to Gleason scores 8-10[25]. The steps to calculate and binarily categorize the prognostic scores were repeated for CSS as given in the previous section for BCR using the same cox regression model for BCR. Kaplan-Meier-Curve was applied to illustrate the cancer-specific survival after stratification according to the low- and high-risk groups.

## Prognostic Evaluation

The coefficient of significance for each variable of the cox regression was estimated using the Wald test. The global statistical significance of the cox regression model was evaluated using the likelihood-ratio test. The LR-test was applied to determine the best fitted model for BCR. We used the log-rank test to find significant differences between Kaplan Meier Curves. One-thousand bootstrap resamples were applied to reduce over-fitting bias and for internal validation. The proportional hazards (PH) assumption of the cox regression model was checked using statistical tests and graphical diagnostics based on the scaled Schoenfeld residuals. The reported p-value is two-sided and statistical significance was defined as $P \leq 0.05$.

## Deep learning models and applied statistical Tools

The detection analyses were based on Python 3.6 (Python Software Foundation, Wilmington, DE) and applied the Keras library which is built-on the TensorFlow framework, to develop the models. All analyses were performed on a GPU machine with 32-core AMD processor with 128 GB RAM (Advanced Micro Devices, Santa Clara, CA), 2 TB PCIe flash memory, 5 TB SDD Hard disks, and a single NVIDIA Titan V GPU with 12 GB VRAM. The statistical analyses were performed using R language and available packages for common statistical approaches and survival analysis (R Foundation for Statistical Computing, Vienna, Austria).

## Acknowledgment

PlexusNet and the derivate digital markers for survival and genomic alteration are patented.